# Decentralized Trusted Timestamping using the Crypto Currency Bitcoin

Bela Gipp, Norman Meuschke, André Gernandt
National Institute of Informatics Tokyo, Japan

**Abstract**
Trusted timestamping is a process for proving that certain information existed at a given point in time. This paper presents a trusted timestamping concept and its implementation in form of a web-based service that uses the decentralized Bitcoin block chain to store anonymous, tamper-proof timestamps for digital content. The service allows users to hash files, such as text, photos or videos, and store the created hashes in the Bitcoin block chain. Users can then retrieve and verify the timestamps that have been committed to the block chain. The non-commercial service enables anyone, e.g., researchers, authors, journalists, students, or artists, to prove that they were in possession of certain information at a given point in time. Common use cases include proving that a contract has been signed, a photo taken, a video recorded, or a task completed prior to a certain date. All procedures maintain complete privacy of the user's data.

**Keywords: Trusted Timestamping, Crypto Currencies, Bitcoin, Block chain, Cryptography, Web Applications**
**Citation**: Editor will add citation
**Copyright**: Copyright is held by the authors.
**Acknowledgements**:
**Contact**: Bela@Gipp.com, N@Meuschke.org, Gernandt@Kontrollfeld.de

## 1   Introduction

Proving the date of content creation is important in many situations and crucial for data used as evidence. For example, inventors must routinely prove when exactly they put forward a patentable invention to obtain or defend a patent. The idea of timestamping information to verifiably establish a time at which it existed is not new. Even before the digital age, information could be encoded, e.g. as an anagram, and the code could be published using a reliably dateable medium, e.g. a newspaper. For digital data, trusted timestamping protocols, which rely on asymmetric cryptography, are used to prove that data has existed and has remained unaltered since a certain point in time (Haber and Stornetta, 1991, Schneier, 1996).

This paper describes a decentralized trusted timestamping concept and its implementation in a web service using crypto currencies to authenticate data integrity. The concept offers the following benefits over established timestamping protocols:

- decentralized, cryptographic integrity validation of the timestamping process;
- high incentives for computing nodes to contribute to the decentralized process;
- minimal effort for users: no need to setup specialized hardware or software;
- low cost of operation, which allows us to provide the service free of charge.

To illustrate the concept, we use a bank transfer as an analogy. Imagine putting a hash value of data, as the reference text of a bank transfer. The time at which the bank performs the transfer and the hash are recorded in the bank's ledger and can serve as a verifiable timestamp as long as the ledger exists. The timestamping concept presented in this paper employs transactions in a crypto currency instead of bank transfers. The transactions are recorded in the block chain of the crypto currency, i.e. the cryptographically validated record of all transactions. In contrast to a bank's ledger, the block chain is not bound to a single entity, but redundantly maintained by all computing nodes participating in the crypto currency network and publically accessible to anyone.

The described timestamping concept is applicable to any crypto currency. With OriginStamp (www.originstamp.org[1]) we provide a non-commercial service that implements the presented concept using the Bitcoin currency. We chose Bitcoin (Nakamoto, 2009), because it has achieved the largest market capitalization and has attracted the highest number of participating computing nodes. We consider market capitalization as an important indicator for the security of a crypto currency. As long as a crypto currency has a value significantly larger than zero, we assume that no possibility to manipulate transactions in the block chain and thereby timestamps exists. The number of participating nodes is an important security characteristic, because the block chain can be manipulated if x>50% of nodes cooperate in acting maliciously (Nakamoto, 2009).

---
[1] Before registering originstamp.org, the service was available at http://gipp.com/originstamp

## 2   Trusted Timestamping

Trusted timestamping processes are specified in RFC 361 (Adams et al., 2001) and the ANSI ASC X9.95 standard (American National Standards Institute (ANSI), 2005), which augmented RFC 361 with data-level security requirements. Both specifications describe processes that require a central time stamping authority (TSA) to issue timestamps and ensure their validity. Figure 1 illustrates the process.

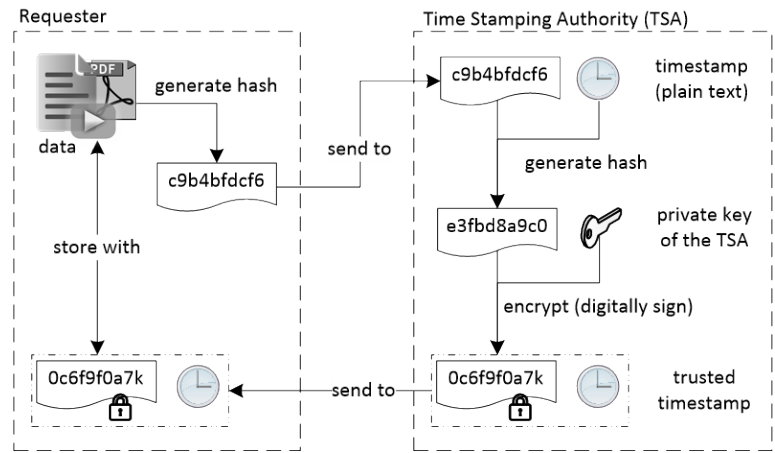

Figure 1: Concept of issuing a trusted timestamp

Initially, the original data is hashed. Hashing authenticates the exact data content, because the hash function ensures that changing a single bit in the data would generate a different hash value. The hash is then transmitted to a TSA, which joins the hash with a plain text timestamp. The resulting string, i.e. the hash combined with the timestamp, is hashed once more and digitally signed using the TSA's private key. The resulting ciphertext represents the trusted timestamp, which, together with the plain text timestamp, is returned to the requester. The validity of the trusted timestamp can be verified by decoding the ciphertext using the public key of the TSA. To verify that some data is identical to the data authenticated by the TSA, the process of creating the trusted timestamp has to be replicated and the results have to be compared to the decoded trusted timestamp.

The need for a central TSA is a weakness of established timestamping approaches, since the integrity of the timestamping process is inevitably bound to the integrity of the TSA (Adams et al., 2001). If the TSA is compromised, all issued timestamps can become invalid given the TSA employs public key encryption using a single private key. Linked encryption or transient key encryption schemes mitigate this weakness (American National Standards Institute (ANSI), 2005, Une, 2001), yet they are technically more demanding and hence involve higher costs of operation. Protocols for decentralized trusted timestamping, in which multiple parties validate timestamps, significantly increase security (Haber and Stornetta, 1991). However, decentralized approaches are technically demanding and require independent parties to cooperate without offering direct incentives to the involved parties for performing their services. Decentralized timestamping approaches have not achieved widespread adoption in practice.

## 3   Decentralized Trusted Timestamping using Crypto Currencies

Crypto currencies can serve as decentralized trusted timestamping services if hash values of digital data are embedded into the transactions recorded in the block chain of the crypto currency. We do not know whether we were the first who had this idea. Shortly after we initially discussed the approach, discussions in bulletin boards started[2,3] and another year later a paper was published (Clark and Essex, 2012).

---

[2] https://bitcointalk.org/index.php?topic=2137.0 (date: 2010-12-07)
[3] https://bitcointalk.org/index.php?topic=52715.0 (date: 2011-11-23)

Figure 2 illustrates the concept using OriginStamp, a service we developed, and the Bitcoin block chain.

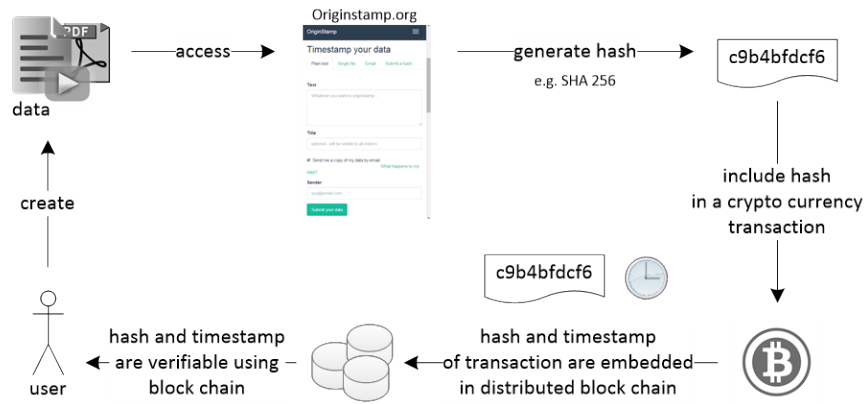

Figure 2: The concept of trusted timestamping using crypto currencies

Figure 3 illustrates the decentralized trusted timestamping process implemented in OriginStamp. The service allows users to timestamp digital content submitted through a web browser, by email, or through plugins to popular applications like WordPress and git. When a user submits a file or plain text through the browser, a client-side Java-Script hashes the data. Alternatively, the data can be hashed offline, e.g., using an open source tool. In either case only the hash, not the data is transmitted to the server. Since the email submission method requires transferring non-hashed data, it is not recommended for highly confidential content because of the risk of interception.

To reduce operating costs, the server collects submitted hashes, concatenates the hashes, and generates a single aggregated hash. By performing additional hashing and encoding operations, the aggregated hash is converted into a Bitcoin address[4]. To this new address, the smallest transactionable amount of Bitcoins (0.00000001 BTC) is transferred. Each transaction requires a fee of currently approx. 0.0001 BTC (3 US cents). By including only the aggregated hash in one Bitcoin transaction per 24 hours, the total transaction costs are less than 10 USD per year. Users are given the option to include their hash in a transaction that is performed immediately for a fee of 1 USD. To provide non-paying users with immediate evidence, their hash is published to Twitter right after submission.

By performing the transaction, the aggregated hash and the timestamp of the transaction are permanently embedded in the distributed Bitcoin block chain. The Bitcoin currency stores transactions as the leaf nodes of a merkle tree (Merkle, 1988). Transactions are formed into a block when a computing node succeeds in finding a number (nonce) that, when inserted into the block, causes the hash of the block to fulfill certain complexity criteria. Among other information, a block contains the root of the merkle tree and the hash of the preceding block, thus forming a chain of blocks. Each of the currently approx. 6,500 computing nodes in the Bitcoin network constantly works on forming blocks, a process referred to as mining. Since each block must reference the preceding block, forming new blocks confirms the content of older blocks. Nodes have an incentive to participate in finding blocks, since a reward is offered for forming a block. Altering the timestamp of a transaction is impossible once a block has been confirmed by five or more subsequent blocks, which requires one hour on average (Nakamoto, 2009).

The timestamp of confirmed transactions and the data they encode can be verified using the OriginStamp website or by inspecting the Bitcoin block chain, e.g. using websites like blockchain.info. The service exclusively stores which hashes were included in which transaction. This information allows verifying any hash using the block chain. Users can choose to receive an email containing the same information to verify their data independent of the OriginStamp service. Therefore, data and timestamps remain verifiable as long as the block chain exits.

---

[4] For details on the conversion process see:
 https://en.bitcoin.it/wiki/Technical_background_of_Bitcoin_addresses#How_to_create_Bitcoin_Address

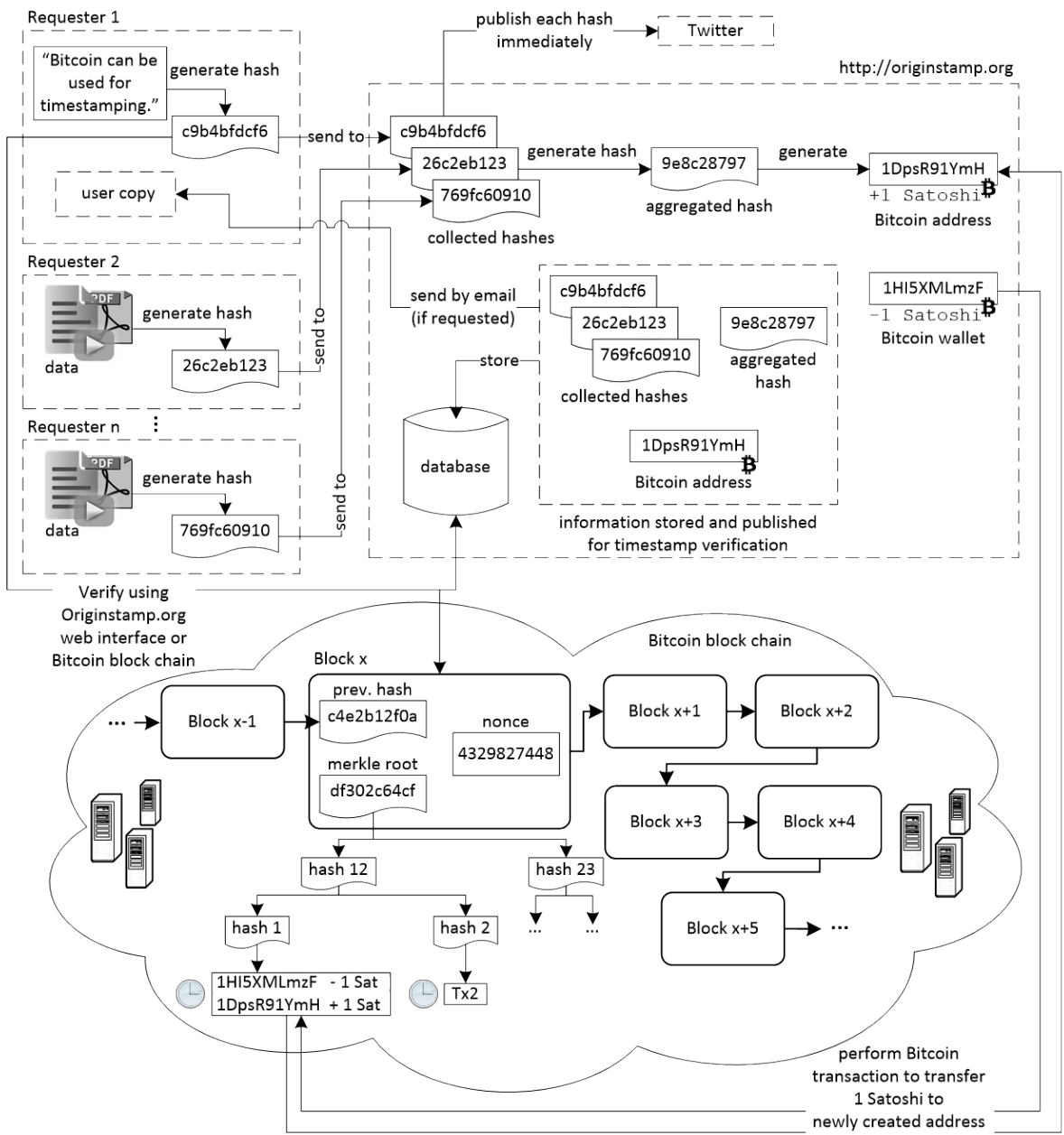

Figure 3: Decentralized trusted timestamping as implemented in Originstamp.org

## 4 Conclusion and Future Work

We presented a service that enables users to anonymously create persistent, tamper-proof timestamps without relying on a centralized timestamping authority. The service presented is non-commercial and uses decentralized trusted timestamping enabled by the distributed and cryptographically validated block chain of the digital currency Bitcoin. In the future, we will develop additional plugins to popular applications to simplify and speed up the creation of timestamps for content of those applications. The timestamping service offers an API capable of creating timestamps in bulk. We invite other parties, including pre-printing services, and the community to use the API to integrate the creation of trusted timestamps as part of their applications and services.

Timestamp of this document[5].

---
[5] http://www.originstamp.org/u/q7hZZglQ3c

# Citation for this Paper

**Citation Example:**

B. Gipp, N. Meuschke, and A. Gernandt. Decentralized Trusted Timestamping using the Crypto Currency Bitcoin. In *Proceedings of the iConference 2015 (to appear)*, Newport Beach, CA, USA, Mar. 24 - 27, 2015. URL http://ischools.org/the-iconference/.

**Bibliographic Data:**

| RIS Format | BibTeX Format |
|---|---|
| TY  - CPAPER<br><br>AU  - Gipp, Bela<br><br>AU  - Meuschke, Norman<br><br>AU  - Gernandt, Andre<br><br>T1  - Decentralized Trusted Timestamping using the Crypto Currency Bitcoin<br><br>T2  - Proceedings of the iConference 2015 (to appear)<br><br>AD  - Newport Beach, CA, USA<br><br>Y1  - 2015/mar. 24 - 27<br><br>UR  - http://ischools.org/the-iconference/ | @INPROCEEDINGS{Gipp15a,<br><br>author = {Gipp, Bela and Meuschke, Norman and Gernandt, Andre},<br><br>title = {Decentralized Trusted Timestamping using the Crypto Currency Bitcoin},<br><br>booktitle = {Proceedings of the iConference 2015 (to appear)},<br><br>year = {2015},<br><br>address = {Newport Beach, CA, USA},<br><br>month = mar # { 24 - 27,},<br><br>url = {http://ischools.org/the-iconference/}<br><br>} |